# Cross-shaped nanostructures for the study of spin to charge interconversions using spin-orbit coupling in non-magnetic materials

V. T. Pham[1], L. Vila[1,*], G. Zahnd[1], P. Noël[1], A. Marty[1], and J. P. Attané[1,†]

[1] SPINTEC, CEA-INAC/CNRS/Univ. Grenoble Alpes, F-38000 Grenoble, France

ABSTRACT

Several spin-orbit effects allow performing spin to charge inter-conversion: the spin Hall effects, the Rashba effect, or the spin-momentum locking in topological insulators. Here we focus on how the detection of this inter-conversion can be made electrically, using three different cross-shaped nanostructures. We apply these measurement configurations to the case of the spin Hall effect in Pt, using CoFe electrodes to detect and inject spins. Both the direct and inverse spin Hall effect can be detected, with a spin Hall signal up to two order of magnitude higher than that of nonlocal measurements in metallic lateral spin valves, and with a much simpler fabrication protocol. We compare the respective signal amplitude of the three proposed geometries. Finally, comparison of the observed spin signals with finite element method calculations allows extracting the spin Hall angle and the spin diffusion length of Pt.

TEXT

Spin current manipulation is at the core of spintronics. Recently, spin-orbit interaction (SOI) in nonmagnetic materials has been exploited to create and detect the spin currents. For instance, the spin Hall effect (SHE) and its inverse [1] allows performing the inter-conversion between charge and spin currents in bulk metals [2]. Edelstein Rashba effects, at Rashba interfaces or at the surfaces of topological insulators, can also be highly competitive for ensuring such interconversions [3].

Direct SHE (DSHE) allows converting a charge current into a transverse pure spin current (PSC), a flow of spin angular momentum without net charge flow. The relationship between the produced PSC ($J_s$), the charge current ($J_c$) and the spin polarization of electrons (**s**) can be written $J_S = \frac{\hbar}{e} \theta_{SHE} J_C \times s$ [1], where the conversion rate $\theta_{SHE}$ is the spin Hall angle of the considered material. Reciprocally, a transverse charge current can be generated from a spin current in the reverse mechanism with the same conversion rate.

---

[*] laurent.vila@cea.fr
[†] jean-philippe.attane@cea.fr

A lot of techniques have been devoted to SHE detection, such as the FM-based spin pumping/spin torque [4, 5, 6], spin Hall magnetoresistance [7], optical schemes [8, 9] and electrical techniques on nanodevices [10, 19, 16]. The nanostructures which allow detecting electrically the SHE are of prime interest in view of SO-based logic devices [11] and non-volatile MRAMs [12], altogether with applications of Spin Orbit torque experiments [13, 14].

Among electrical methods, non-local techniques which exploit abilities of ferromagnets for injecting and detecting spin current in lateral spin valves (LSV) [15, 16] is playing a significant role in SHE researches. Notwithstanding, the complex nanostructure fabrication and the small absolute values of the spin signals, especially for metallic systems, is hampering its straightforward use.

Electrical devices which can produce high spin signals are thus of prime importance not only for basics research but also for applications. Recently, a simple F/N bilayer device with two ferromagnetic electrodes probing directly the spin accumulation induced by the SHE of an underneath metallic nanowire with strong SOC has been proposed to achieve large spin-charge inter-conversion signals [17]. In this paper, we compare DSHE and ISHE measurement on different cross-shaped nanostructures. Thanks to the local spin detection technique, our experiments show mΩ-DSHE/ISHE signals, up to two orders of magnitude larger compared to classical non local devices, with a much simpler structure that could be implemented into SHE-based computational devices [11, 12]. The extracted spin Hall angle of Pt in these CoFe/Pt systems by the FEM simulations shows a good agreement with the spin Hall angle of Pt extracted in these Co/Fe Systems by FEM simulations [*17*].

The samples have been fabricated by conventional e-beam lithography, e-gun deposition, and lift-off processes on the thermal oxidized $SiO_2$/Si substrate. The chosen materials are $Co_{60}Fe_{40}$ and Pt, as they are archetypal ferromagnetic and SHE materials. Their magnetic and spin transport properties have been previously characterized [17, 18]. The depositions were performed in high vacuum, about 4e-8 mTorr. The top surface of the Pt nanowires is cleaned by Ar ion etching prior to the deposition of the CoFe electrode. They are connected by Au/Ti electrical pads. In all the samples the thickness of the CoFe layer is 15 nm, whereas that of Pt is 7 nm for the device of Fig. *1(a)*, and 8 nm for the samples of *Figs. 1(c)* and *1(e)*. The widths of the CoFe and Pt nanowires are 50 nm and 400 nm, respectively. The transport measurements were performed using a lock-in amplifier working at 332 Hz, with an applied current of 100 or 200 µA. In the presented measurement configurations, the spin signal in Ohm is the ratio between the detected voltage and the applied current. In all experiments, the field is applied along the easy axis of the ferromagnetic electrodes. *Figures 1(a), 1(c) and 1(e)* show the SEM images and the measurement principle of the studied devices in view of DSHE configuration, whereas ISHE measurement setups are sketched on *Figs. 1(b, d, f)*.

*Figures 1(a)* and *1(b)* represent the device proposed in *ref. 17*. In *Fig. 1(a),* a vertical PSC is created by DSHE when flowing an electrical current along the Pt wire. The PSC is probed by the two CoFe electrodes, providing that the magnetizations of the two ferromagnetic-electrodes are opposite. Otherwise they probe the same electrochemical potential and no voltage difference develops between them. Reciprocally in *Fig. 1(b),* a non-zero net spin polarized current is injected vertically at the two interfaces of CoFe/Pt only when the two magnetizations of the CoFe electrodes are opposite. The ISHE of Pt wire converts this spin current into a transverse charge current which can be detected as a voltage in open circuit condition.

A simplified design of a cross-shaped device is shown in *Figs. 1(c)* and *1(d)*, with only a single CoFe electrode on top of a T-shaped pattern made of Pt. In *Fig. 1(c)*, the charge current flowing in the Pt creates a spin accumulation at its top surface. The transverse contacts probe the electrochemical potential associated to the spin accumulation at the CoFe/Pt top interface. The polarization of the ferromagnetic electrode controlled by an external magnetic field, defines if majority or minority spins are probed. The inverse SHE is measured by permuting the current and voltage leads [*Fig. 1(d)*]. The charge current flowing in the CoFe allows injecting a spin current into the Pt stripe. This spin current is converted into a transverse charge current by ISHE, which leads to the development of a transverse voltage in open circuit conditions. The direction of the produced charge current is reversed when the polarization of the spin current changes with the reversal of the magnetization of the ferromagnetic electrode. *Figures 1(e)* and *1(f)* illustrate a simple cross made of two straight CoFe and Pt wires, respectively designed for ISHE and DSHE measurements using the same principles.

The magnetoresistance loops of the DSHE and the ISHE are plotted in *Figs. 2. (a, c, and e) and Figs. 2. (b, d, and f)* respectively, using the configurations shown in *Figure 1*. All the figures have the same range of signal amplitude to ease their direct comparison, an arrow on Fig. 2(c) depicts a 5 mΩ amplitude. The drop and rise of the transverse resistance corresponds to the magnetization reversal of the CoFe contact, leading to the change of the probed electrochemical potential (majority or minority) in DSHE, or to the change of sign of injected spin current polarization in ISHE. The amplitude of the spin signals is the difference between the different plateaus.

These simple nanostructures show that the absolute value of the spin-to-charge signal is in the 10 mΩ range at room temperature, around two orders of magnitude larger than standard nonlocal measurement technique in metallic LSVs [15, 19, 20]. For instance, the signal is even comparable to the tunnelling spin Hall signals obtained in LSVs having a 2D-graphene channel that owns superior spin transport properties [21, 22]. The enhancement of the spin signals in our device takes also advantage of the stronger spin current injection of CoFe compare to NiFe [18].

As expected for ISHE and DSHE, the obtained signals have similar amplitude when changing the measurement configuration, as expected by their reciprocal relation. We believe that the small offset resistance can be explained by the slight misalignment between the different electrodes, as seen in the SEM image in *Fig. 1(d)*.

The results illustrated in *Figs. 2(a) and 2(b)* correspond for the measurement setups carried out on device geometry similar to that of *ref. 17*. The total spin signal, the difference of resistance level for the two anti-parallel states, is of the order of 16 mΩ. The amplitude of the signal is larger by nearly a factor of 2 than the 7 mΩ for the device shown in Figs. *2 (c) and 2(d)*. Indeed, in the measurements of Fig. 2(a-b) the spin current is injected twice at the F/NM interface in ISHE or the spin accumulation measured twice in DSHE, the signal is thus roughly a factor of 2 larger in the sample geometry presented in *Figs. 1(a-b)* with respect to that of *Figs. 1(c-d)*. Nonetheless, the geometry of *Figs.1 (c-d)* has larger tolerance in the alignment error between the ferromagnetic wires and the Pt slab, and might thus be easier to scale down.

Figures *2(e)* and *2(f)* illustrate the magnetoresistance loops corresponding to the geometry device and probe configuration of *Figs. 1(e)* and *1(f)*, respectively. A zero-spin signal is obtained in both cases. In the DSHE configuration, the same electrochemical potential is probed along the ferromagnetic wire when it is homogenously magnetized and thus no voltage develops [Fig. 2(e)]. In the ISHE configuration, the injected spin current into the Pt wire is zero in average, thus producing no transverse voltage. We will show later how to access the spin signal in this device.

As shown in *Figures 3,* these cross nanostructures allow studying the spin signal using other probe configurations, as depicted in the insets, with the voltage and current leads circulating by pairs around the sample. This type of measurement has been used in experiments involving Topological Insulators [23, 24, 25]. The first observation is that Figs. 3(a) and 3(b) combines DSHE and ISHE signals. They can be distinguished by the two switching events and the plateau on each branch of the hysteresis, corresponding to the achievement of the antiparallel magnetic states of the CoFe electrodes. Since a nucleation pad is patterned at the end of the bottom ferromagnetic electrode (F1), it has a lower reversal field. Depending on this contact being either injecting a current or detecting a voltage, the corresponding reversal event (at low absolute field value) corresponds to either ISHE or DSHE signal. This is the opposite for the reversal event at larger absolute field. The second observation is that the spin signal, around 7 mΩ, is roughly two times lower than that of *Figs. 2(a-b)*. Indeed, now only one detection (DSHE) and injection (ISHE) mechanism occurs here.

The same behaviour is observed in the T-shape device. The spin signals in *Fig. 3(c)* and (d) are roughly a factor of 2 smaller than that of Figs. 2c and 2d, 3 mΩ compared to 7 mΩ. The signal is decreased because only half of the transverse voltage is measured in ISHE, or half of the spin accumulation in DSHE.  In contrast the probe configuration of *Figs. 3(e)* and *3(f)* allows now the observation a signal for the device with a straight CoFe wire. The results can be explained by using the same mechanism in *Figs. 2(a)* and *2(b)*, *i.e.* being a combination of DSHE and ISHE. However, there is no distinction between DSHE and ISHE contributions in the signal here because both the injector and detector have the same magnetic state. Additionally, because the ferromagnetic-wire is continuous the transverse resistance is decreased and the spin signal amplitude is much reduced compare to the results in Figs. 3(a) and 3(b), where instead the CoFe wire doesn't fully cover the Pt strip.

Note that the anomalous Hall effect (AHE) [26] can also appear with the same principle as the SHE in all of the measurements. However, the AHE plays an insignificant role (less than 10 %) in those thin samples [see more details in the supplementary material of *ref. 17*].

The device presented in Fig. 1(c) has thus the advantage of showing a much larger signal than that of *Fig. 1(e)*, and a simpler fabrication process than that of *Fig. 1(a)*. In order to estimate the spin-charge conversion rate of this system, we carried out finite element method simulations, similarly as in *ref. 17* for the sample of *Figs. 1(a-b)*, within the framework of a 2 spin-current drift diffusion model [27].

*Figure 4* describes the simulations in ISHE configuration for the scheme shown in *Fig. 1(c)*. The distribution of the applied charge current in *Fig. 4(a)* allows injecting the spin polarized current in Z direction at the CoFe/Pt interface. The contour of the spin accumulation in *Figs. 4(b)* and *4(c)* is for a spin current flowing in the $-Z$ direction when the magnetization is polarized along $+X$. The amplitude of the spin signal is the difference of the two calculated values of the transverse voltage. We retrieved the same spin Hall angle and the spin diffusion length ($\lambda_s$) of Pt reported in *ref. 17 for a* system made of CoFe/Pt, $\Theta_{SHE}$ = 0.19 and $\lambda_s$ = 3.0 nm, with the measured resistivity $\rho_{Pt}$ = *28 µΩcm* for Pt at a thickness of 8 nm. The resistivity of CoFe is $\rho_{CoFe}$ = 20 *µΩcm*, for a thickness of *15 nm*.

To conclude, we studied the spin-charge inter-conversion on simple cross-shaped nanostructures. The milliohm-spin signals are obtained in both the DSHE and ISHE configurations thanks to this local measurement technique. Additionally, we can measure DSHE and ISHE in a simple cross junction between CoFe and Pt wires. The FEM simulations were performed to analyse the spin Hall conversion rate. Apart from giving a straightforward technique for the metrology of SHE/ISHE, these results are also promising for developing computational magnetic devices, such as a non-volatile memory with a single nano-magnet [12] or spin-orbit based logic circuits [11]. These device geometries can also be used to study the spin-charge interconversion using Rashba interfaces or Topological Insulators.

*Figure Captions:*

**Figure 1:** *(a), (c),* and *(e)* show the SEM image of the studied devices and the schemes for ISHE measurements. In *(a)* and *(c)*, the ferromagnetic electrodes, CoFe (in dark-grey color) are patterned on top of the transverse nanowire and of a T-shaped stripe made of Pt (in wine color). An applied charge current **Jc** injects a spin current **Js** from the CoFe electrode into the Pt layer along – Z-direction at the interfaces. Thanks to the ISHE in the Pt layer, the spin current is deflected orthogonally to the plane defined by the spin **S** (X) of electron and **Js** (Z), i.e., a transverse charge current is generated along the Y direction, which is a Y–Z cross-section of the device. A voltmeter is used to detect the ISHE voltage ($V_{ISHE}$) in the nanowire along the Y-direction. *(e)* The opposite spin current in each side of the interface expects a negligible spin signal. The external magnetic field (yellow arrow) controls the magnetization in the FM electrode, represented by the black arrows. *(b), (d),* and *(f)* are sketches of the devices showing the direct SHE mechanism and measurement setup, reciprocal to *(a), (c),* and *(e)*.

**Figure 2:** *(a) – (f)* are the loops recorded in direct or inverse spin Hall effect experiments corresponding to the measurement schemes shown in Figs. 1(a) – 1(f), respectively. All the curves are have the same size and an arrow on (c) shows a 5 m$\Omega$ signal amplitude. The black arrows represent the magnetization states. The dashed arrows denote the direction of the applied magnetic field.

**Figure 3:** Alternative measurement configurations for all three proposed devices**.** *(a)* and *(b)* present the recorded loop for the sample of Figure 1(a,b), (c) and (d) corresponds to the device of Figure 1(c,d) and that of (e) and (f) to that of Figure 1 (e,f). The corresponding measurement schemes are shown in the inset of the figures.

**Figure 4:** *(a)* Charge current lines calculated by finite element method simulations for the sample of Fig 1 (d) and showing the spread of the current line in the Pt strip. A charge current is applied along the ferromagnetic electrode through the Pt wire in X direction of the T-shape pattern. *(b)* and *(c)* are lateral cut-planes (X-Y cut-plane) of the electrochemical potential contour in the case of + X-magnetization, at the interface (z=0) and at a distance z = – 7.5 nm. The magnetization directions are represented by the bright arrows.

Acknowledgements

The devices were fabricated in the Platforme Technologie Amont in Grenoble. We acknowledge the support from the labex laboratory LANEF of Univ. Grenoble Alpes, and funding from the ANR TOPRISE.

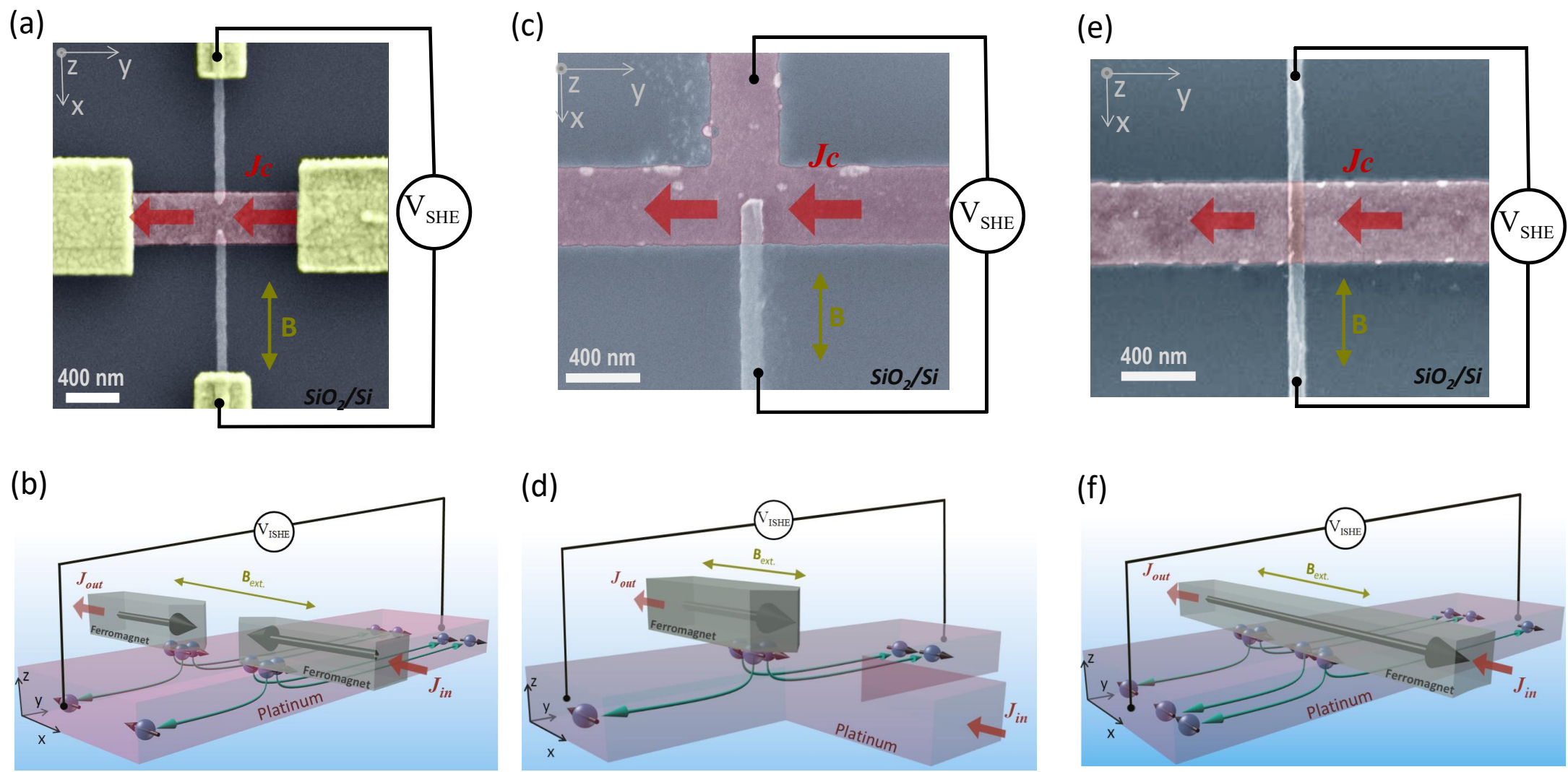

Figs. 1(a), (b), (c), (d), (e) and (f)

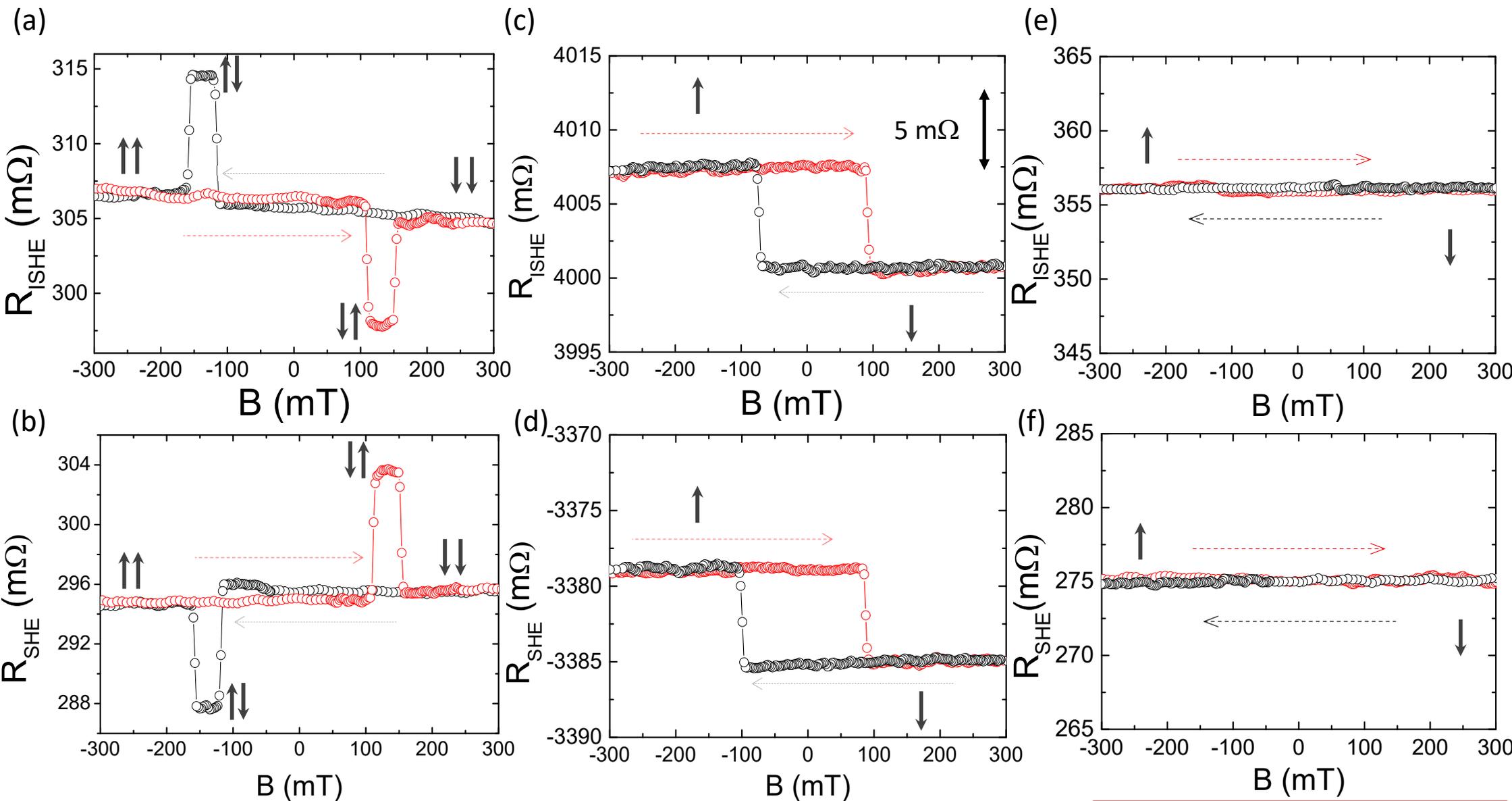

Figs. 2 (a), (b), (c), (d), (e) and (f)

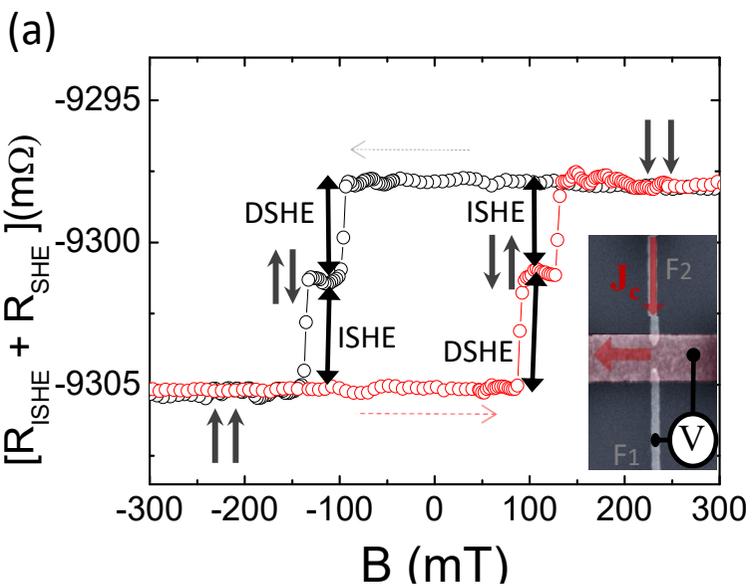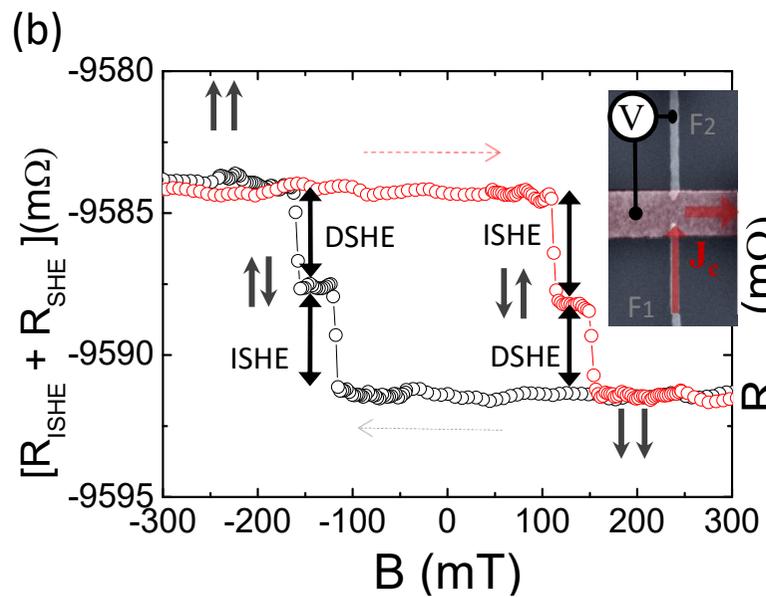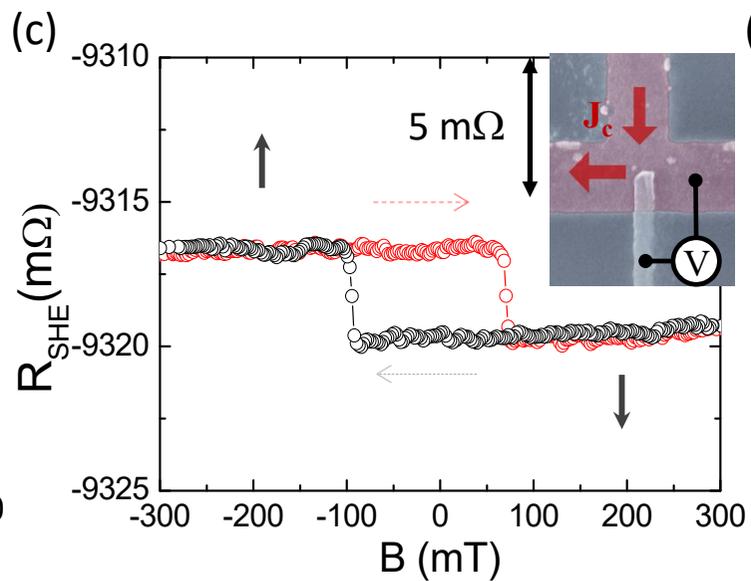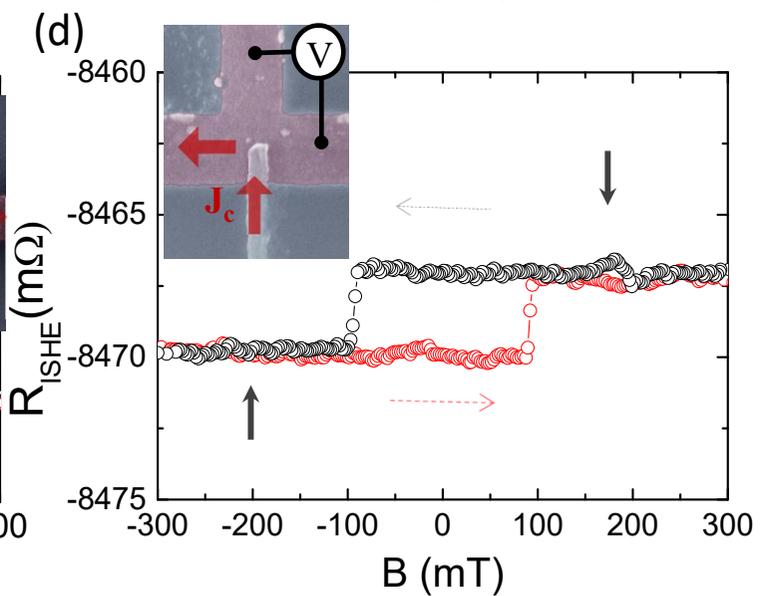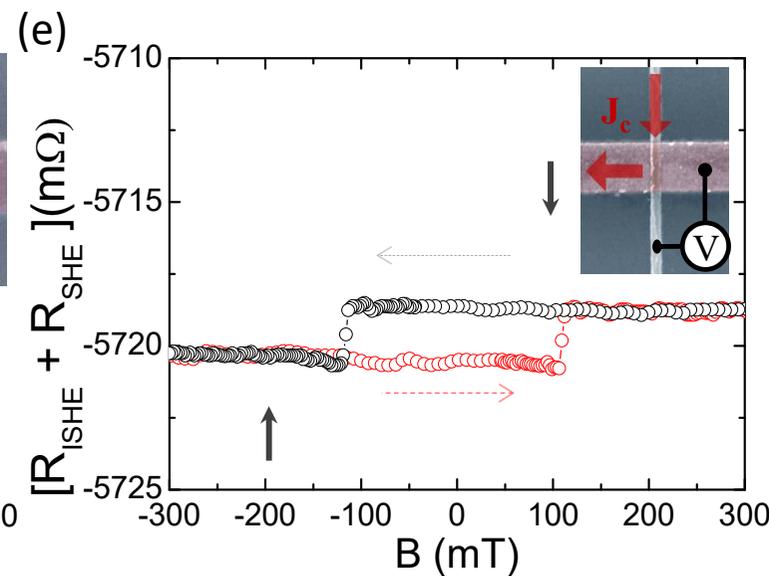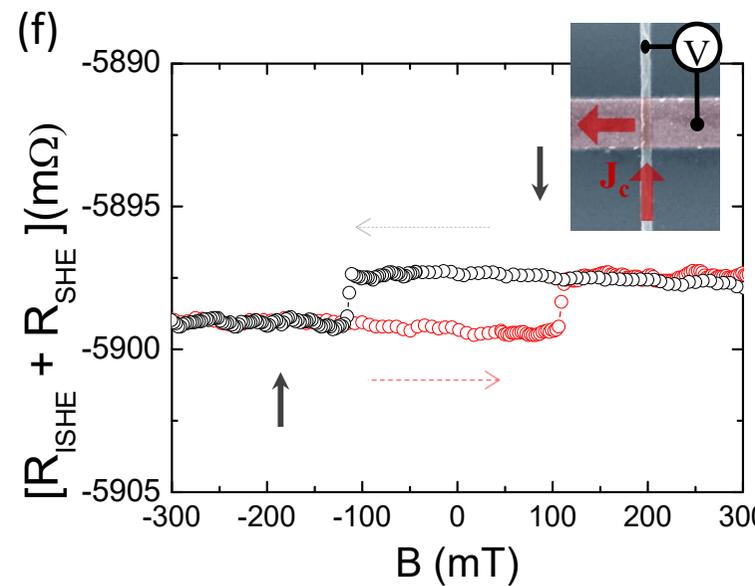

Figs. 3 (a), (b), (c), (d), (e) and (f)

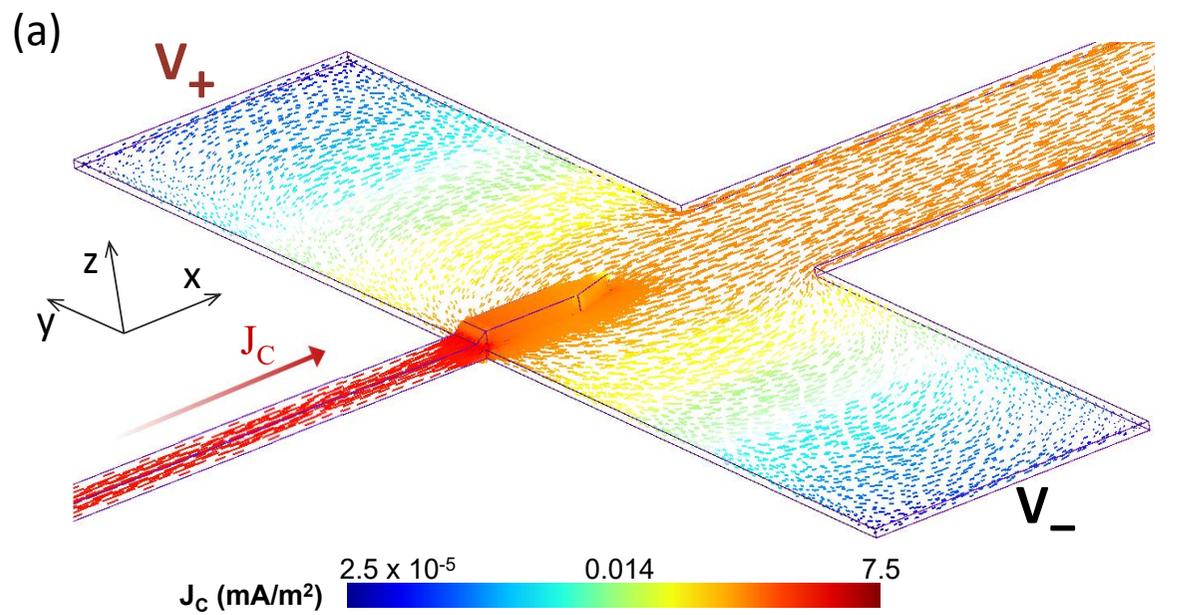
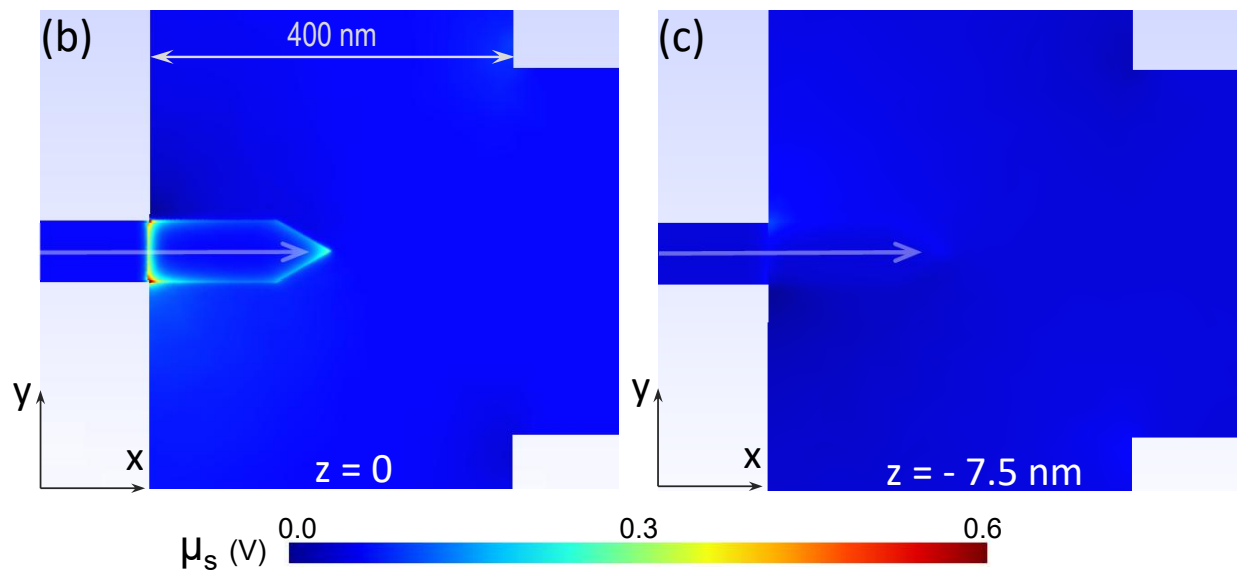

Figs. 4 (a), (b) and (c)